# DEPLOYING CONTAINERIZED QUANTEX QUANTUM SIMULATION SOFTWARE ON HPC SYSTEMS.

John Brennan∗[1], Momme Allalen†[2], David Brayford†[3], Kenneth Hanley∗[4], Luigi Iapichino†[5], Lee J. O'Riordan∗[6] and Niall Moran∗[7]

∗Irish Centre for High End Computing, National University of Ireland, Galway, Galway, Ireland
†Leibniz Supercomputing Centre of the Bavarian Academy of Sciences and Humanities, Garching b. Muenchen, Germany
[1]john.brennan@ichec.ie, [2]momme.allalen@lrz.de, [3]david.brayford@lrz.de, [4]kenneth.hanley@ichec.ie,
[5]luigi.iapichino@lrz.de, [6]lee.oriordan@ichec.ie, [7]niall.moran@ichec.ie

## ABSTRACT

The simulation of quantum circuits using the tensor network method is very computationally demanding and requires significant High Performance Computing (HPC) resources to find an efficient contraction order and to perform the contraction of the large tensor networks. In addition, the researchers want a workflow that is easy to customize, reproduce and migrate to different HPC systems.

In this paper, we discuss the issues associated with the deployment of the QuantEX quantum computing simulation software within containers on different HPC systems. Also, we compare the performance of the containerized software with the software running on "bare metal".

## 1 INTRODUCTION

As the HPC systems move towards Exascale, the diversity of HPC architectures has increased with new compute unit designs that combine traditional processors with accelerators and memory technologies. These technologies include X86 and AArch64 many core processors, GPUs, FPGAs and ASICS, as well as new memory technologies such as High Bandwidth Memory (HBM) incorporated on the processor die, non-volatile memory technologies that are fast enough to be used as non-volatile RAM (NVRAM) with the storage capacity of a solid-state drive (SSD) for accelerated IO performance.

This increase in processor heterogeneity requires a significant software development effort to ensure that the applications can take advantage of the compute performance available in these new architectures. In addition, the software developer will need to spend effort ensuring that their applications can run on as many different HPC architectures as possible.

The result is that the end users will have significantly more complex configuration and build procedures, which can result in the incorrect building of the applications. This results in the application developers spending more time supporting their users to build the software on specific architectures. Containers provide a potential solution to this problem, by enabling the software developer to provide a correctly configured software for the specific architecture to the end users, either in the form of an image or a recipe in the form of a Dockerfile [1].

There is an increasing interest in the use of HPC resources by non-traditional HPC users who want to analyze massive amounts of data and solve increasing complex "real world" problems. Amongst the most cost-effective solutions is to use pre-existing HPC infrastructures, the possibility of scaling out to more compute nodes and parallelizing the workloads to speed-up the whole process.

However, the deployment of some software and workflows are not straightforward and require a lot of time and effort by the HPC center and scientists to build and maintain the different software that the end users need to conduct their research. Often it is not possible to install the required versions of the software because it requires the system software to be modified, which is often impractical on production HPC systems.

To enable scientists to take advantage of the massive amount of compute power available on large HPC systems, one needs to find a mechanism that enables the scientist to deploy their software and workflows in a way that is simple and does not require a lot of modification to their code and workflow, but also respects the existing HPC system environment, workflows, and security policies. To achieve this, we employ the use of container technologies developed for secure HPC systems at Los Alamos National Laboratory (LANL).

The purpose of this paper is to highlight the work done by the Leibniz Supercomputing Center (LRZ) and the Irish

Centre for High-End Computing (ICHEC) to enable the execution of quantum simulations within containers on various HPC systems, without compromising security, stability, and performance. In addition, we highlight the challenges associated with various containerized software applications and workflows on the different HPC systems at LRZ.

## 2 RELATED WORK

In this section we introduce some commonly used containerized environments on computer systems.

### 2.1 User Defined Software Stack

With the increasing demand on more flexible execution models from researchers the standard methods to offer software on HPC systems (i.e. predefined modules with specific software packages) becomes impractical. User Defined Software Stacks (UDSS) as defined by Priedhorsky et al. [2] combined with the recently introduced user namespaces in Linux offer a solution to how those can be realized in the form of containers without sacrificing security of the HPC cluster.

The UDSS concept helps to deal with problems such as handling dependencies, frequent software updates and support of other Linux environments.

User namespaces available in Linux, since Linux Kernel 3.8, allows the execution of privileged operations without escalating permissions up to root. No child processes do not have control over their parent processes. Mappings for UIDs and GIDs between parent and children processes make sure that operations are executing in a safe way by using the original ID of the user.

#### 2.1.1 Docker

Docker [1] is considered an industry standard container that provides the ability to package and run an application in an isolated environment. However, Docker was not designed for use in a secure multi-user HPC environment and has significant security issues, which enables the user inside the Docker container to have privileged (root) access on the host systems network filesystem, making it unsuitable for HPC systems.

#### 2.1.2 Singularity

Singularity [3] was developed at LBL to be a containerization solution for HPC systems and supports several HPC components such as resource managers, job schedulers and contains built in MPI features. Singularity uses the layered OCI container image and snapshot it into a SIF file.

Although Singularity has been developed to run in a non- privileged namespace, potential security issues came to light during a security review at LRZ in 2018, which have since been resolved in later versions. As a result of the internal security review and concerns of the system administrators. LRZ current policy is not to allow Singularity on SuperMUC-NG.

#### 2.1.3 Shifter

Shifter [4] was developed at NERSC in collaboration with Cray to enable Docker images to be securely executed on an HPC ecosystem. Shifter enables users to transform Docker images to a common format that can then be distributed and launched on HPC systems.

Shifter appears to be a good choice for conventional HPC batch queuing infrastructure. It also provides a scalable and performant solution but retains as much compatibility as possible with the Docker workflow. However, Shifter requires more administrative setup than the equivalent Charliecloud UDSS.

#### 2.1.4 Charliecloud

Charliecloud [2] was developed at LANL to be a lightweight open source UDSS implementation based on the Linux user namespace for HPC sites with strict security requirements and at the time of this work consists of less than 4000 lines of code. It uses Docker to build the HPC container image, shell scripts to unpack the image to an appropriate location and a C program to activate the image and run user code within the image.

In these secure environments, Charliecloud's distinct advantage is the usage of the newly introduced user namespace to support non-privileged launch of containerized applications. The user namespace is an unprivileged namespace and within the user namespace, all other privileged namespaces are created without the requirement of root privileges, which means that a containerized application can be launched without requiring privileged access to the host system.

Charliecloud employs Docker tools locally, but unlike Docker, the container is flattened to a single archive file in preparation for execution. To enable the distributed execution of tasks across multiple nodes, the containerized application is scaled by distributing the archive file to each compute node and unpacked into a local tmpfs environment.

Brayford et al. [5, 6] describe a mechanism that employs Charliecloud to deploy TensorFlow and train a complex neural network at scale on large scale secure HPC production systems.

### 2.1.5 Podman

Podman [7] is an open source container management tool for developing, managing and deploying containers on Linux systems. For HPC, one usually disables network, PID and IPC namespaces. It is considered a mainly unprivileged rootless container and details of the containers privilege are described in depth by Priedhorsky et al. [8].

Podman appears to be a good choice for secure containers as it runs as a rootless and is supported by RedHat. However, the use of Podman has not been fully investigated at LRZ for its security implications, and thus its use has not been allowed yet on our HPC systems. We will consider this container technology for future tests.

### 2.1.6 Container Performance and Overheads

HPC systems come with limited amount of resources, for example, system memory and storage. It is important that newly introduced software stacks, such as the HPC containers, exhibit minimal overhead in terms of runtime and utilization of the system resources [9-12]. For example, the amount of system memory on a single compute node in HPC is limited and exceeding the limit causes the job to fail. A.Torrez et al. [13] describe the performance impacts of HPC container runtimes on HPC benchmarks such as High Performance Linpack (HPL), High Performance Conjugate Gradient (HPCG) and STREAM. Brayford et al. [6] describe the overheads of HPC containers of ML benchmarks such as AlexNet.

Recently, Ruhela et al. [14] have demonstrated the use of HPC container technologies to execute jobs with Petascale performance on numerous CPU and GPU HPC systems, while Brayford et al. [5] have executed distributed machine learning training at Petascale on a large CPU only system.

*2.2 Software*

### 2.2.1 Python

Python [15] is an interpreted high-level programming language, which is used extensively in the fields of pre- and post-processing of data, artificial intelligence and specifically machine learning. The language is widely considered accessible for prototyping algorithms, due to the inherent constructs and the availability of optimized scientific and AI/ML specific libraries, compared to traditional programming languages such as C++. In addition, because Python is an interpreted programming language, it does not need to be recompiled for different CPU architectures unlike C++.

However, a significant issue using Python on secure HPC systems is the way Python is employed within machine learning, often requires a connection to the Internet to install packages. As a result, deploying Python software on a secure HPC system without a connection to the Internet is problematic.

### 2.2.2 Julia

Julia [16] is an open source high level dynamic programming language developed for scientific and numerical computing. It uses an LLVM based JIT compiler to generates native machine code. This enables the user to not only quickly prototype algorithms, but also provide a mechanism to easily achieve performance like that of compiled C/C++ code.

Although most publicly available frameworks for Quantum Computing simulations are based on Python, there is a growing interest in Julia as language for HPC applications Regier et al. [17], which motivates the choice of this language for our project. Another reason for choosing Julia is the ease with which GPUs can be used as described by Besard et al. [18].

One issue that caused a problem when using Julia in a containerized environment was that the default Julia environment and packages directories are located in the user's home directory at ~/.julia . However, by default Charliecloud mounts the host HOME directory into the container image at runtime. So, anything installed in home during the container creation is not visible to runtime. To overcome this, we set the Julia environment path by changing the environment variable JULIA_DEPOT_PATH to /opt/Julia

While Julia provides comparable performance to compiled languages, due to its JIT nature, the first time a code block is run can incur significant time and memory overhead for compilation. For long running computations this time amortized over the duration of the run, but for smaller computations this can be a significant issue. One approach to overcoming this is to use the PackageCompiler package [19] to pre-compile the necessary functions into a custom system image which avoids this compilation at runtime. Another significant benefit of using containers is that these custom system images can be included in the container image which avoids this expensive compilation step at runtime.

### 2.2.3 Like I Knew What I'm Doing. (LIKWID)

LIKWID [20] was designed to address four problems that are frequently encountered by programmers during code migration and code optimization. First, it reports the thread and cache topology of a node so developers can better understand what architecture exactly they are working with. Secondly, it can pin processes and threads to cores used in a program to achieve better thread affinity. Thirdly, LIKWID uses the Linux Model Specific Registers (MSR) module, reads the MSR files from user space and reports the hardware performance counters for several performance metrics. Finally, it provides a set of micro-benchmark kernels to enable developers to quickly test some characteristics and features of the architecture.

We use the LIKWID software to compare the performance profiles of the quantum gate simulation software run on "bare metal" directly on the system with the same run inside the container on the different architectures at LRZ.

To enable LIKWID installed in the hosts module system to be used inside the container, we modify the environment file (/ch/environment) generated during the Docker image creation to contain the environment setting for LIKWID on the host and mount the modules directory inside the container with the bind command when the container was executed using the command in interactive mode:

```
$ ch-run -w -set-env=<environment_file> -b /host/modules/.:/host/modules ./container_image -- bash
$ likwid-perfctr -m -g MEM_DP -C 0 /instrumented_app
```

In batch mode:

```
$ ch-run -w -set-env=<environment_file> -b /host/modules/.:/host/modules ./container_image -- likwid-perfctr -m -g MEM_DP -C 0 /instrumented_app
```

Notice that the mounted host directory after the option -b contains . before the : and the corresponding container directory name. This was added to ensure that the host directory binds correctly inside the container at runtime even if it is a symbolic link.

### 2.3 Quantum Computing

### 2.3.1 Efficient Quantum Circuit Simulation on Exascale Systems (QuantEX)

Quantum computing technology is just emerging from its infancy: Whereas experiments were previously limited to systems with only a handful of qubits, the first Noisy Intermediate Scale Quantum Systems (NISQ) are now emerging, which feature upwards of 50 qubits. However, quantum hardware is still very experimental, and its limitations are such, that algorithmic development must resort to simulations running on traditional platforms to make progress. Because of the algorithmic complexity and the memory footprint, simulating quantum algorithms has become a typical HPC challenge. The QuantEX projects aims to develop simulation tools that require less compute and storage capacity but can still simulate the operation of NISQ devices. Tensor network contractions, mathematical methods developed by quantum information theorists and condensed matter physicists for evaluating systems of interacting particles, are well suited to this goal.

Within the project, we have been able to successfully convert the containerized quantum simulation applications to a Charliecloud HPC container and have successfully executed the benchmarks on SuperMUC-NG (SNG) production system and the BEAST test systems.

### 2.4 QXContexts

QXContexts [21] is a Julia package for executing simulations of quantum circuits using a tensor network approach and targeting large, distributed memory clusters with hardware accelerators. It was developed as part of the QuantEx project, one of the individual software projects of work package 8 (WP8) of PRACE 6$^{th}$ implementation phase (6IP).

To simulate a quantum circuit, an associated package called QXTools is used to define the simulation as an efficient sequence of tensor contractions and prepare a set of simulations files outlining the simulation using a domain specific language. QXContexts is designed to parse these simulation files and perform the tensor contractions making use of distributed compute resources via MPI as well as hardware accelerators.

## 3 SYSTEM & COMPONENTS

### 3.1 SuperMUC-NG

The SuperMUC-NG (SNG) system [22] at the Leibniz Supercomputing Center of the Bavarian Academy of Science and Humanities (BADW-LRZ), has a peak performance of 26.9 petaflops, which consists of 311,040 Intel® Xeon® Platinum 8174 (Skylake-SP) CPU cores and 719 terabytes of main memory. The Skylake cores are arranged into eight "thin" and one "fat" island. The "thin" islands consist of 6336 two-socket nodes with each node containing 48 CPU cores and equipped with 96 GB of main memory.

The systems software stack is OpenHPC [23] compliant and the main development environment is Intel Parallel Studio XE 2019 & Intel MPI 2019.

For security reasons, SNG has no direct connection to the Internet on both the login and compute nodes, and SSH has been disabled on the compute nodes.

### 3.2 Bavarian Energy, Architecture, and Software Testbed (BEAST)

The Bavarian Energy Architecture and Software Testbed (BEAST) is a collection of small test systems for the research and evaluation of new hardware technologies. Currently BEAST consists of three different CPU architectures: AMD X86, Arm ThunderX2 and Arm Fujitsu A64fx.

The AMD systems consists of two node Rome GPU 2U servers, with two AMD EPYC 7742 with 64 cores along with 512GB of DDR4-3200, two 1.9 Terabyte SSD and two AMD Radeon MI-50 GPUs with 32 Gigabytes of high bandwidth memory (HBM). The interconnections between the nodes are Mellanox InfiniBand: HDR 200Gb/s.

The ARM ThunderX2 (THX2) system consists of two node GPU servers each consisting of two 32 core Arm ThunderX2 CPUs with 512 gigabytes of memory and two Nvidia V-100 GPUs with 32 gigabytes of HBM. That are connected with Mellanox EDR InfiniBand 100Gb/s.

Finally, The Fujitsu A64fx system is an eight node HPE system consisting of Arm Fujitsu A64fx CPUs with 64 cores and two 512 bit vector units and 32 gigabytes HBM2 memory. That is connected with a Mellanox InfiniBand EDR interconnect.

### 3.3 Charliecloud Conversion

To be able to deploy Docker images on SNG they first must be converted into a Charliecloud container. Before we can proceed, Docker and Charliecloud must be installed on a system, where they have root access. Also, we discovered that Charliecloud commands to convert the Docker image to a Charliecloud image require privileged (root) access on the systems we have been using due to a bug in the tar software application used [24].

When installing software in the containerized image one should keep in mind any restriction in the access of external servers. For example, the command 'pip install' will not succeed on SNG because the system does not allow access via https.

In addition, software should not be installed in $HOME within the container, because by default Charliecloud mounts the users $HOME from the host system into the container. During the execution of the Charliecloud container, files that the user has installed in $HOME within the Docker image are no longer accessible, even though the files are visible outside the Charliecloud run time environment on the host system. This is a particular problem for the Julia programming environment and we described this issue in more detail in the Julia section.

## 4 RESULTS

In this section, we discuss how we deployed the containerized software on the different CPU architecture systems at LRZ, compare the ease of deployment and execution time with that of the same software installed and executed on "bare metal". The Julia software was executed three times and within each execution the simulation was performed twice and we present the measurements of the median execution time of the Julia software application.

### 4.1 BEAST Systems Segments

As the BEAST experimental systems are only available for internal purposes and we did not have access to other Arm based systems to execute privileged operations, it was decided to create the Docker images and convert them to Charliecloud images on the experimental HPC systems.

#### 4.1.1 Arm THX2 CPU Host

The first step was to install Julia, QXContexts and LIKWID on the THX2 system, which was achieved by following the standard installation instructions for the software packages.

To be able to profile the benchmark simulation with LIKWID one first had to create a Julia file called likwid.jl. That provides an interface between the Julia simulation code and the LIKWID profiler, which we included as a header file in the file qxrun.jl. We then instrumented the specific simulation routines that we were interested in with the LIKWID function prototypes we defined inside the file likwid.jl. This was achieved by enclosing the call to the simulation function within a likwid_markerStartRegion and likwid_markerStopRegion function call as illustrated in figure 1. It has been decided to profile the L2 cache miss rate/ratio, which is calculated from the CPU hardware counters, because we consider this group of performance diagnostics an extremely relevant performance comparison across multiple architectures for the tensor network contraction operations. The tensor contraction is essentially a generalization for matrix multiplication and it is well known that cache useage has a big impact on the performance of linear algebra operations on large matrices. The timing information as shown in table 1 for the compilation and execution of the simulation code was performed using a timing function within the Julia language as shown in figure 1.

```
reset_timer!()
  results = execute(…, timings=timings)
if timings
 likwid_markerInit()
  ret_val = likwid_markerRegisterRegion("SIM")
  reset_timer!()
   likwid_markerStartRegion("SIM")
     results = execute(…, timings=timings)
   likwid_markerStopRegion("SIM")
 likwid_markerClose()
```

**Figure 1:** **Timing and profiling code example**

When QXContexts is executed, the first step involves the setup, build and execution of the Julia application. So, to avoid having the setup and compilation times included in the performance measurements and application profiling, we call the simulation function twice and measure separately the time taken for first and second execution of the simulation using the Julia timing function. However, we only instrument the second call to the simulation function with LIKWID.

| Tot / % measured: | Time | Allocations |
|---|---|---|
| Compile + exec | 47.4s / 96.8% | 3.79GiB / 98.4% |
| Exec only | 185ms / 41.3% | 11.7MiB / 79.2% |

**Table 1:** **QXContexts compilation and execution time.**

Notice the significant difference in the execution times between the first time the simulation function is executed and second time the simulation executes as shown in table 1. This is because during the first execution of the simulation, Julia compiles the code using the LLVM JIT compiler before executing the simulation. This means that the measured execution time includes both the compilation time and execution time, while during the subsequent execution of the simulation only the native execution of the simulation occurs.

The results of the LIKWID profiling of the second time the simulation is executed provides information from the hardware counters for L2 cache miss ratio of the CPU are shown in table 2.

| RDTSC Runtime [s] | | 0.185925 |
|---|---|---|
| call count | | 1 |
| Event | Counter | HWThread 0 |
| INST_RETIRED | PMC0 | 444784700 |
| CPU_CYCLES | PMC1 | 448009600 |
| L2D_CACHE | PMC2 | 26456010 |
| L2D_CACHE_REFILL | PMC3 | 5377605 |
| Metric | | HWThread 0 |
| Runtime (RDTSC) [s] | | 0.1859 |
| Clock [MHz] | | 2409.6254 |
| CPI | | 1.0073 |
| L2 request rate | | 0.0595 |
| L2 miss rate | | 0.0121 |
| L2 miss ratio | | 0.2033 |

**Table 2:** **LIKWID L2CACHE report**

### 4.1.2  Arm THX2 CPU Container

In our next experiment, we run the same application on the THX2 CPU, but within a Charliecloud container. Please note that the combined setup, compilation and simulation times between the containerized version in table 3 and the bare metal version in table 1 are very similar. Similarly, the performance of the containerized version of the simulation and the bare metal version are almost identical.

| Tot / % measured: | Time | Allocations |
|---|---|---|
| Compile + exec | 51.1s / 97.0% | 3.93GiB / 98.5% |
| Exec only | 184ms / 41.1% | 12.0MiB / 79.7% |

**Table 3:** **QXContexts compilation and execution time**

The results of the LIKWID L2 cache diagnostics from the hardware performance counters shown in table 4 are very

Deploying Containerized QuantEX Quantum Simulation Software on HPC Systems.

similar within a few percent to the results of the non-containerized version in table 3. This is what we would expect as the difference in execution times between the bare metal and containerized version is negligible.

| RDTSC Runtime [s] | | 0.187872 |
|---|---|---|
| call count | | 1 |
| Event | Counter | Event |
| INST_RETIRED | PMC0 | 446985100 |
| CPU_CYCLES | PMC1 | 452133800 |
| L2D_CACHE | PMC2 | 27119410 |
| L2D_CACHE_REFILL | PMC3 | 5679842 |
| Metric | | HWThread 0 |
| Runtime (RDTSC) [s] | | 0.1879 |
| Clock [MHz] | | 2406.6056 |
| CPI | | 1.0115 |
| L2 request rate | | 0.0607 |
| L2 miss rate | | 0.2094 |
| L2 miss ratio | | 0.2094 |

**Table 4:** LIKWID L2CACHE report

### 4.1.3 Arm Fujitsu A64fx Host

Table 5 is analogous to table 1, but for the A64fx system.

| Tot / % measured: | Time | Allocations |
|---|---|---|
| Compile + exec | 104s / 97.0% | 3.79GiB / 98.4% |
| Exec only | 414ms / 44.7% | 11.7MiB / 79.2% |

**Table 5:** QXContexts compilation and execution time on A64fx

For the A64fx architecture the L2 cache miss ratio (L2CACHE) performance metrics were not available in the LIKWID tool at the time we performed the experiments, so we decided to look at the arithmetic and memory performance metrics (MEM_DP) LIKWID. In table 6, we show the metrics for the bare metal execution.

| Region Info | | HWThread 0 |
|---|---|---|
| RDTSC Runtime [s] | | 0.433888 |
| call count | | 1 |
| Event | Counter | HWThread 0 |
| FP_DP_FIXED_OPS_SPEC | PMC0 | 496 |
| FP_DP_SCALE_OPS_SPEC | PMC1 | 0 |
| L2D_CACHE_REFILL | PMC2 | 220698 |
| L2D_CACHE_WB | PMC3 | 74772 |
| L2D_SWAP_DM | PMC4 | 16940 |
| L2D_CACHE_MIBMCH_PRF | PMC5 | 24380 |
| Metric | | HWThread 0 |
| Runtime (RDTSC) [s] | | 0.4339 |
| DP (FP) [MFLOP/s] | | 0.0011 |
| DP (FP+SVE128) [MFLOP/s] | | 0.0011 |
| DP (FP+SVE256) [MFLOP/s] | | 0.0011 |
| DP (FP+SVE512) [MFLOP/s] | | 0.0011 |
| Memory read bandwidth [MBytes/s] | | 105.8355 |
| Memory read data volume [GBytes] | | 0.0459 |
| Memory write bandwidth [MBytes/s] | | 44.1165 |
| Memory write data volume [GBytes] | | 0.0191 |
| Memory bandwidth [MBytes/s] | | 149.9521 |
| Memory data volume [GBytes] | | 0.0651 |
| Operational intensity (FP) | | 7.623451e-06 |
| Operational intensity (FP+SVE128) | | 7.623451e-06 |
| Operational intensity (FP+SVE256) | | 7.623451e-06 |
| Operational intensity (FP+SVE512) | | 7.623451e-06 |

**Table 6:** LIKWID memory arithmetic report

### 4.1.4 Arm Fujitsu A64fx Container

Comparing the execution times between table 5 and table 7, we notice that the containerized version is ~3 seconds faster to build and execute the first simulation than the bare metal version. However, as this difference is relatively small, we can consider the results to be the same.

| Tot / % measured: | Time | Allocations |
|---|---|---|
| Compile + exec | 101s / 96.9% | 3.68GiB / 98.4% |
| Exec only | 457ms / 50.0% | 11.7MiB / 79.2% |

**Table 7:** QXContexts compilation and execution time

When comparing the execution times of the second simulation, we notice that the bare metal version is slightly faster, again this difference is relatively small, and the difference could be accounted for by a system task running in the background.

Table 8 depicts the arithmetic and memory performance metrics provided by the LIKWID tool. When comparing the results with those shown in table 9, we notice that the runtime difference is consistent with the results reported by the application.

| Region Info | | HWThread 0 |
|---|---|---|
| RDTSC Runtime [s] | | 0.477314 |
| call count | | 1 |
| Event | Counter | HWThread 0 |

| FP_DP_FIXED_OPS_SPEC | PMC0 | 496 |
|---|---|---|
| FP_DP_SCALE_OPS_SPEC | PMC1 | 0 |
| L2D_CACHE_REFILL | PMC2 | 883647 |
| L2D_CACHE_WB | PMC3 | 373491 |
| L2D_SWAP_DM | PMC4 | 48554 |
| L2D_CACHE_MIBMCH_PRF | PMC5 | 40377 |
| Metric | | HWThread 0 |
| Runtime (RDTSC) [s] | | 0.4773 |
| DP (FP) [MFLOP/s] | | 0.0010 |
| DP (FP+SVE128) [MFLOP/s] | | 0.0010 |
| DP (FP+SVE256) [MFLOP/s] | | 0.0010 |
| DP (FP+SVE512) [MFLOP/s] | | 0.0010 |
| Memory read bandwidth [MBytes/s] | | 426.2337 |
| Memory read data volume [GBytes] | | 0.2034 |
| Memory write bandwidth [MBytes/s] | | 200.3161 |
| Memory write data volume [GBytes] | | 0.0956 |
| Memory bandwidth [MBytes/s] | | 626.5498 |
| Memory data volume [GBytes] | | 0.2991 |
| Operational intensity (FP) | | 1.658525e-06 |
| Operational intensity (FP+SVE128) | | 1.658525e-06 |
| Operational intensity (FP+SVE256) | | 1.658525e-06 |
| Operational intensity (FP+SVE512) | | 1.658525e-06 |

**Table 8:** LIKWID memory arithmetic report

When comparing the memory bandwidths reported by LIKWID between the bare metal and container runs, we notice that the memory read bandwidth was ~4 times faster, the memory write data bandwidth was ~4.5 times faster and the memory bandwidth was ~4.2 faster for the containerized execution. However, the arithmetic intensity was ~4.6 times faster for the bare metal execution. Further investigation is required to better understand these reported differences.

### 4.1.5 AMD Rome CPU Host

On the AMD Rome nodes we decided not to profile the application using LIKWID due to time constraints of configuring and installing the necessary software. However, we did execute QXContexts on bare metal and within a container. The execution times are shown in tables 9 and 10.

| Tot / % measured: | Time | Allocations |
|---|---|---|
| Compile + exec | 24.9s / 96.9% | 3.67GiB / 98.4% |
| Exec only | 96.7ms / 41.2% | 11.7MiB / 79.2% |

**Table 9:** QXContexts compilation and execution time

### 4.1.6 AMD Rome CPU Container

Comparing the run times of the compilation and first simulation between the bare metal version shown in table 9 and the containerized version in table 10. We can report that the times are identical

| Tot / % measured: | Time | Allocations |
|---|---|---|
| Compile + exec | 24.9s / 96.8% | 3.68GiB / 98.3% |
| Exec only | 39.3ms / 100% | 9.30MiB / 100% |

**Table 10:** QXContexts compilation and execution time

However, when we compare the results of the second simulation, we notice that the containerized execution was faster. This result needs further investigation to fully understand why the containerized version was significantly faster than the bare metal version.

### 4.2 Productions System SNG

As a last step of our investigation, the same software run on the BEAST segments has been tested on SuperMUC-NG, the flagship-class HPC system at LRZ, to verify that the software can be deployed on a more restricted system

### 4.2.1 Intel X86 Host

Table 11 reports the same information as table 1, but for the Intel Skylake system.

| Tot / % measured: | Time | Allocations |
|---|---|---|
| Compile + exec | 28.0s / 97.0% | 3.69GiB / 98.4% |
| Exec only | 192ms / 100% | 9.31MiB / 100% |

**Table 11:** QXContexts compilation and execution time

The results of the LIKWID profiling of the hardware counters for L2 cache miss ratio (L2CACHE) are shown in table 12.

| Region Info | HWThread 0 | |
|---|---|---|
| RDTSC Runtime [s] | 0.193949 | |
| call count | 1 | |
| Event | Counter | HWThread 0 |
| INSTR_RETIRED_ANY | FIXC0 | 205929200 |
| CPU_CLK_UNHALTED_CORE | FIXC1 | 108736200 |
| CPU_CLK_UNHALTED_REF | FIXC2 | 99488740 |
| L2_TRANS_ALL_REQUESTS | PMC0 | 6640719 |
| L2_RQSTS_MISS | PMC1 | 453140 |
| Metric | HWThread 0 | |
| Runtime (RDTSC) [s] | 0.1939 | |
| Runtime unhalted [s] | 0.0454 | |
| Clock [MHz] | 2616.8545 | |

| CPI | 0.5280 |
|---|---|
| L2 request rate | 0.0322 |
| L2 miss rate | 0.0022 |
| L2 miss ratio | 0.0682 |

**Table 12:** LIKWID L2CACHE report

### 4.2.2 Intel X86 Container

Comparing the results of the compilation and first simulation run of the bare metal version from table 11 with the containerized results in table 13. We notice that the execution time of the containerized version is faster.

| Tot / % measured: | Time | Allocations |
|---|---|---|
| Compile + exec | 27.8s / 96.9% | 3.68GiB / 98.3% |
| Exec only | 118ms / 100% | 9.30MiB / 100% |

**Table 13:** QXContexts compilation and execution time

However, when comparing the results of the second simulation that the containerized version is again significantly faster than that of the bare metal.

When we compare the L2 cache miss ratio reports in tables 12 & 14, we notice that the CPU clock frequency is ~119MHz faster, the cycles per instruction (CPI) value is 0.1384 greater, the L2 request rate is 0.0281 larger, the L2 miss rate is 0.0071 greater and the L2 miss ratio is 0.0867 larger for the containerized version that for the bare metal execution.

| Region Info | HWThread 0 | |
|---|---|---|
| RDTSC Runtime [s] | 0.119785 | |
| call count | 1 | |
| Event | Counter | HWThread 0 |
| INSTR_RETIRED_ANY | FIXC0 | 247875700 |
| CPU_CLK_UNHALTED_CORE | FIXC1 | 165182600 |
| CPU_CLK_UNHALTED_REF | FIXC2 | 144563300 |
| L2_TRANS_ALL_REQUESTS | PMC0 | 14940460 |
| L2_RQSTS_MISS | PMC1 | 2314073 |
| Metric | HWThread 0 | |
| Runtime (RDTSC) [s] | 0.1198 | |
| Runtime unhalted [s] | 0.0690 | |
| Clock [MHz] | 2735.8267 | |
| CPI | 0.6664 | |
| L2 request rate | 0.0603 | |
| L2 miss rate | 0.0093 | |
| L2 miss ratio | 0.1549 | |

**Table 14:** LIKWID L2CACHE report

We speculate that the higher CPU clock frequency of the run within the container is the reason for the faster execution time of that run. Also, it is possible that the node of the bare metal simulation was "hot" from a previous job run and the TDP limit gets hit and the CPU automatically clock the frequency down. It is not trivial to relate this difference to the variation of the other L2 diagnostics, and we prefer to postpone a more detailed analysis of this point to future work.

## 5 CONCLUSIONS

The results of this work have been extremely promising and justify the effort we have made in developing the QuantEX software with HPC specific containers for HPC systems with different architectures and operational modes.
We have shown that it is relatively simple to build, configure and deploy the quantum gate simulation software developed in Julia on multiple different HPC systems that contained different CPU architectures and instruction sets. We were also able to show that for most cases the run time performance difference for QXContexts between containerized and bare metal versions to be negligible. There is a tendency for containerized versions to be marginally faster than bare metal version in some systems, and at least for Intel X86 we could trace it back to the clock frequency, although the root cause of it is unclear. It is left for future work to understand if the same occurs also for the other architectures who showed faster execution time in the containerized version.

Also, we have demonstrated that it is possible to combine the software environments of the container and host system, to enable the host installed LIKWID software to be used to profile the application within the container.
Finally, we expect that the use of HPC containers will continue to increase, as more users with non-traditional workflows need high-end HPC resources.

## 6 FUTURE WORK

This work is to be intended as proof of concept for deploying quantum gate simulation software written in Julia as containers on different HPC systems with different CPU architectures, and we have started to extend this to different HPC GPU architectures. On the technical side, we would also like to extend this work to investigate the behaviour of the LIKWID profiler measuring the hardware MSR counters from inside a container environment. In addition, we would like to repeat the experiments with a higher number of simulations to be able to perform some statistical analysis on the measurement.

## ACKNOWLEDGMENTS

The authors gratefully acknowledge the Gauss Centre for Supercomputing e.V. (*www.gauss-centre.eu*) for funding this project by providing computing time on the GCS Supercomputer SuperMUC at Leibniz Supercomputing Centre (*www.lrz.de*). This work was financially supported by the PRACE project funded in part by the EU's Horizon 2020 Research and Innovation programme (2014-2020) under grant agreement 823767.

Deploying Containerized QuantEX Quantum Simulation Software on HPC Systems.